\documentclass[cits]{PoS}

\title{Strangeness in the nucleon from a mixed action calculation}

\ShortTitle{Strangeness in the nucleon from a mixed action calculation}

\author{\speaker{Michael Engelhardt}%
\\
        Department of Physics, New Mexico State University, Las Cruces,
        NM 88003, USA\\
        E-mail: \email{engel@nmsu.edu}}

\abstract{The contributions of strange quarks to the nucleon mass and
the nucleon spin are investigated in a mixed action scheme employing
domain wall valence quarks and quark loops on MILC asqtad dynamical
fermion ensembles. Results are presented for pion masses 495 MeV and
356 MeV.}

\FullConference{The XXVIII International Symposium on Lattice Field Theory\\
                 June 14-19,2010\\
                 Villasimius, Sardinia, Italy}

\begin{document}

\section{Introduction}
Strange quarks provide a special perspective on nucleon structure due
to their absence from the valence content of the nucleon. As a result
of this absence, effects of quark-antiquark pair creation and annihilation
can be studied in isolation. Correspondingly, considerable efforts to probe
strange quarks in the nucleon have been, and continue to be made on the
experimental side. On the one hand, electron scattering experiments have
provided data on the strange electric and magnetic form factors \cite{emff},
as well as given some indication of the strange axial structure; the latter,
on the other hand, can also be studied in more detail in neutrino scattering
experiments, with a combined analysis having been presented in \cite{pate}.
Considerably enhanced data are expected to emerge from the upcoming neutrino
experiments MicroBooNE and MINER$\nu$A.

On the side of lattice QCD, calculating strange contributions to nucleon
structure counts among the relatively hard problems, since these contributions
are determined purely by disconnected diagrams, requiring propagator
traces, cf.~Fig.~\ref{ddiagram}. On the other hand, among this hard
class of problems, strange matrix elements are still the easiest to
access, since the quark being propagated in the disconnected loop is
heavy, and the associated propagator calculation is therefore less
expensive than for light quark loops. Thus, besides the physical
relevance of such calculations, they can also serve as initial test
cases for exploring techniques to evaluate disconnected contributions
to hadron physics more generally. A number of such investigations are
being pursued by a variety of groups \cite{bc1,bc2,freeman,doi,bab,jlq1,jlq2}.
The effort presented here focuses on the two most fundamental contributions
of strange quarks to the properties of the nucleon, namely, their
contributions to the nucleon mass and the nucleon spin. The calculational
scheme employed is one which has been developed and optimized by the
LHP Collaboration \cite{lhpc10}. Domain-wall quark propagators are
evaluated on HYP-smeared MILC asqtad dynamical quark ensembles. While
the use of domain wall fermions implies considerable computational 
expense, it is expected to yield advantages in terms of mild
renormalization and chiral behavior.

\section{Strange matrix elements}
\begin{figure}
\begin{center}
\includegraphics[width=6cm]{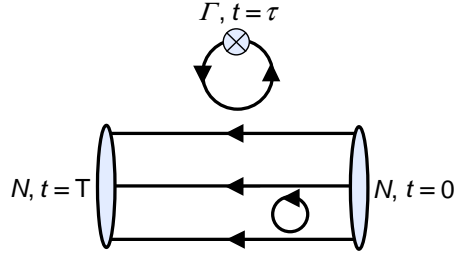}
\end{center}
\vspace{-1.2cm}
\caption{Disconnected contribution to nucleon matrix elements. The nucleon
propagates between a source at $t=0$ and a sink at $t=T$; the
insertion of $\Gamma \equiv \Gamma^{obs} $ occurs at an intermediate time
$t=\tau $.}
\label{ddiagram}
\end{figure}

The strange contributions to nucleon mass and spin can be characterized by
the matrix elements
\begin{equation}
f_{T_s } =\frac{m_s }{m_N } \langle N | \bar{s} s | N\rangle
\ \ \ \ \ \ \ \ \ \ \ \ \ \mbox{and}
\ \ \ \ \ \ \ \ \ \ \ \ \ 
\Delta s = \langle N,i | \bar{s} \gamma_{i} \gamma_{5} s
| N,i \rangle
\label{matel}
\end{equation}
respectively, where $| N,i \rangle $ denotes a nucleon state with spin
polarized in the $i$ direction. These matrix elements are obtained from
corresponding lattice correlator ratios,
\begin{equation}
R[\ \Gamma^{nuc} , \Gamma^{obs} \ ] (\tau ,T)
= \frac{\left\langle \ \left[
\Gamma^{nuc}_{\alpha \beta } \ \Sigma_{\vec{x} } \
N_{\beta } (\vec{x} ,T) \bar{N}_{\alpha } (0,0) \right]
\cdot \left[ - \Gamma^{obs}_{\gamma \rho }
\Sigma_{\vec{y} } \ s_{\rho } (\vec{y} ,\tau )
\bar{s}_{\gamma } (\vec{y} ,\tau ) \right] \
\right\rangle }{\left\langle \
\Gamma^{unpol}_{\alpha \beta } \ \Sigma_{\vec{x} } \
N_{\beta } (\vec{x} ,T) \bar{N}_{\alpha } (0,0) \ \right\rangle }
\label{corrrat}
\end{equation}
where $\bar{N} , N$ denote (smeared) nucleon sources and sinks, the sums
over spatial position $\vec{x} $ project onto zero momentum nucleon states,
the standard minus sign accompanying the quark loop has already become
explicit through the reordering of the strange quark fields,
and the $\Gamma $ matrices implement nucleon polarization and operator
insertion structure. Specifically, the unpolarized nucleon two-point
function in the denominator of (\ref{corrrat}) is achieved using
$\Gamma^{unpol} = (1+\gamma_{4} )/2 $; on the other hand, the three-point
function in the numerator results from evaluating the correlation between
a nucleon propagator and a strange quark loop, as also displayed
diagrammatically in Fig.~\ref{ddiagram}. To obtain $f_{T_s } $, one
calculates
\begin{equation}
\frac{m_s }{m_N } \left( R[\ \Gamma^{unpol} , 1\ ] (\tau ,T)
\ - \ [\mbox{VEV}] \right)
\ \equiv \ R\{ f_{T_s } \} \
\longrightarrow \ f_{T_s }
\label{corrfts}
\end{equation}
in the limit $T \gg \tau \gg 0$, where, as indicated, the vacuum
expectation value of the quark loop, $[VEV] = \langle -\Sigma_{\vec{y} } \
s_{\gamma } (\vec{y} ,\tau ) \bar{s}_{\gamma } (\vec{y} ,\tau ) \rangle $
is subtracted; the matrix elements (\ref{matel}) are meant to characterize
the strange content of the nucleon {\em relative} to the vacuum, which itself
contains a strange scalar condensate. The nucleon mass $m_N $ can be
extracted from the nucleon two-point function as a by-product of the
calculation.
On the other hand, to obtain $\Delta s$, one calculates
\begin{equation}
-i\, \cdot 2 \cdot R[\ (-i\gamma_{i} \gamma_{5} /2) \ \Gamma^{unpol} ,
\gamma_{i} \gamma_{5} \ ] (\tau ,T)
\ \equiv \ R\{ \Delta s \} \
\longrightarrow \ \Delta s
\label{corrds}
\end{equation}
in the limit $T \gg \tau \gg 0$. In this case, the corresponding
vacuum expectation value vanishes, but it can nevertheless be calculationally
advantageous to subtract this numerical zero in order to reduce statistical
fluctuations. Note that, in (\ref{corrds}), an average has already been
taken over the expectation values obtained using nucleons polarized in the
positive and the negative $i$-directions, respectively; averaging the
corresponding projectors $(1\mp i\gamma_{i} \gamma_{5} )/2$ (with a relative
minus sign) leads to the first argument of $R$ in (\ref{corrds}). In the
numerical calculation, also the polarization axis $i$ will be averaged over
the three spatial directions in order to further improve statistics, cf.~the
description further below. The prefactor 2 in (\ref{corrds}) is a normalization
factor compensating for the fact that the unpolarized nucleon two-point
function is always used in the denominator of the ratio (\ref{corrrat}),
even when polarized nucleon states are used in the numerator. Finally, the
prefactor $(-i)$ incorporates the Wick rotation back to Minkowski space;
it compensates for the additional factor $i$ arising in the $\gamma_{5} $ 
matrix contained in the second argument of $R$ in (\ref{corrds}) when
casting the calculation on the Euclidean lattice.

\section{Lattice setup}
\begin{figure}
\vspace{-0.46cm}
\begin{center}
\includegraphics[angle=-90,width=11.1cm]{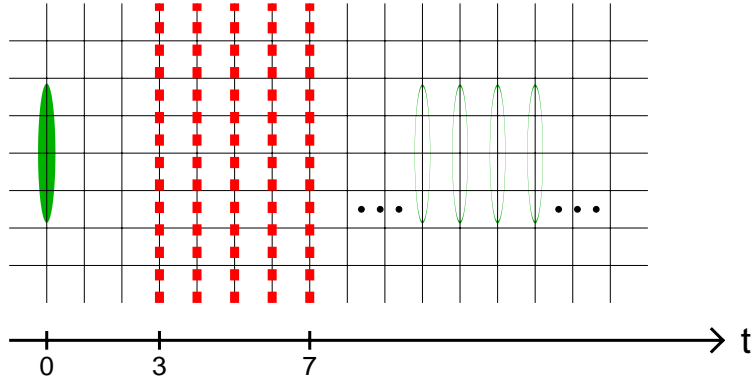}
\end{center}
\vspace{-1.1cm}
\caption{Setup of the lattice calculation. The nucleon source is located
at lattice time $t=0$. The operator insertion time $\tau $ is averaged
over $t=3,\ldots ,7$; accordingly, stochastic sources are distributed
over the bulk of the lattice in this entire time range. The temporal
position $T$ of the nucleon sink is variable.}
\label{setup}
\end{figure}

Numerical work was carried out on two $2+1$-flavor asqtad dynamical quark
ensembles provided by the MILC collaboration, corresponding to the pion masses
$m_{\pi } = 356\, \mbox{MeV} $ and $m_{\pi } = 495\, \mbox{MeV} $. These
ensembles contained $448$ and $486$ $20^3 \times 64$ lattices, respectively,
with lattice spacing $a=0.124\, \mbox{fm} $. The configurations were
HYP-smeared for the purpose of this calculation. The nucleon two-point
functions and strange quark loops were evaluated using domain wall quarks.
The lattice setup employed is depicted in Fig.~\ref{setup}. To enhance
statistics, the operator insertion time $\tau $ in (\ref{corrrat}) was
averaged over five time slices, $t=3,\ldots ,7$ (where $t=0$ corresponds to
the nucleon source position). To implement this average, complex $Z(2)$
stochastic sources, introduced to the evaluate the strange quark loop
propagator trace, were distributed over the bulk of the lattice within this
entire temporal range. For $m_{\pi } = 356\, \mbox{MeV} $, 1200 stochastic
sources per configuration were used, for $m_{\pi } = 495\, \mbox{MeV} $,
600 stochastic sources\footnote{These high statistics in the stochastic
sources mainly serve to improve the signal for $\Delta s$; the stochastic
estimator for the scalar matrix element, by contrast, converges rapidly
and the statistical uncertainty in $f_{T_s } $ is dominated by gauge
fluctuations.}. This rigid temporal setup, chosen to allow for maximal
statistics, does not permit a variation of the operator insertion time in
order to test for a plateau; it is motivated by previous extensive experience
with connected diagrams in the same scheme \cite{lhpc10}, which suggests that
the nucleon ground state has been filtered out at $t=3$ to a sufficient
degree as to render the associated systematic uncertainty small compared to
the statistical uncertainty of the present calculation. On the other hand, a
residual opportunity to test the dependence of the results specifically on
the operator-sink separation is given; below, results for the relevant
correlator ratios will be shown as a function of variable sink position $T$,
with the expectation that the asymptotic behavior will be reached for sink
positions $T\ge 10$.

Besides statistical uncertainties stemming from the stochastic evaluation of
the quark loop, the observables studied here exhibit substantial gauge
fluctuations. To accumulate statistics sufficient to overcome these
fluctuations, it is necessary to evaluate multiple samples of the correlator
ratios of interest per given gauge configuration. Thus, for given source
time slice, not one, but several (eight in the case of
$m_{\pi } = 356\, \mbox{MeV} $, four in the case of
$m_{\pi } = 495\, \mbox{MeV} $) different samples were obtained by varying
the spatial source position. Furthermore, since the scheme depicted in
Fig.~\ref{setup} requires a much smaller temporal extent than available on the
lattice employed, the entire scheme was replicated three times on separate
temporal regions of the lattice. Altogether, therefore, 24 correlator ratio
samples were obtained per lattice gauge configuration in the case of
$m_{\pi } = 356\, \mbox{MeV} $, and 12 in the case of
$m_{\pi } = 495\, \mbox{MeV} $. Finally, as already indicated in the
previous section, $\Delta s$ was averaged over three separate polarization
axes $i$, corresponding to the coordinate axes.

\section{Numerical results and conclusions}
\begin{figure}
\vspace{-0.3cm}
\includegraphics[angle=-90,width=7cm]{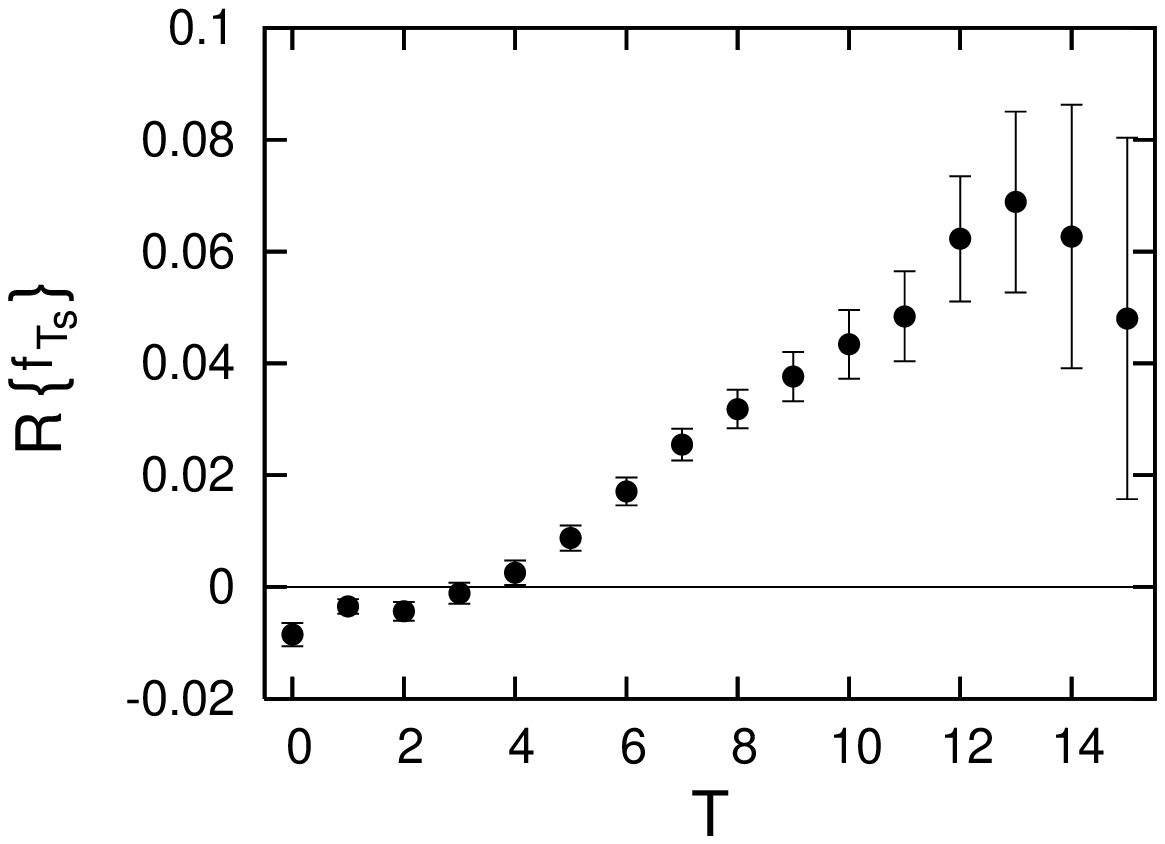}
\hspace{0.95cm}
\includegraphics[angle=-90,width=7cm]{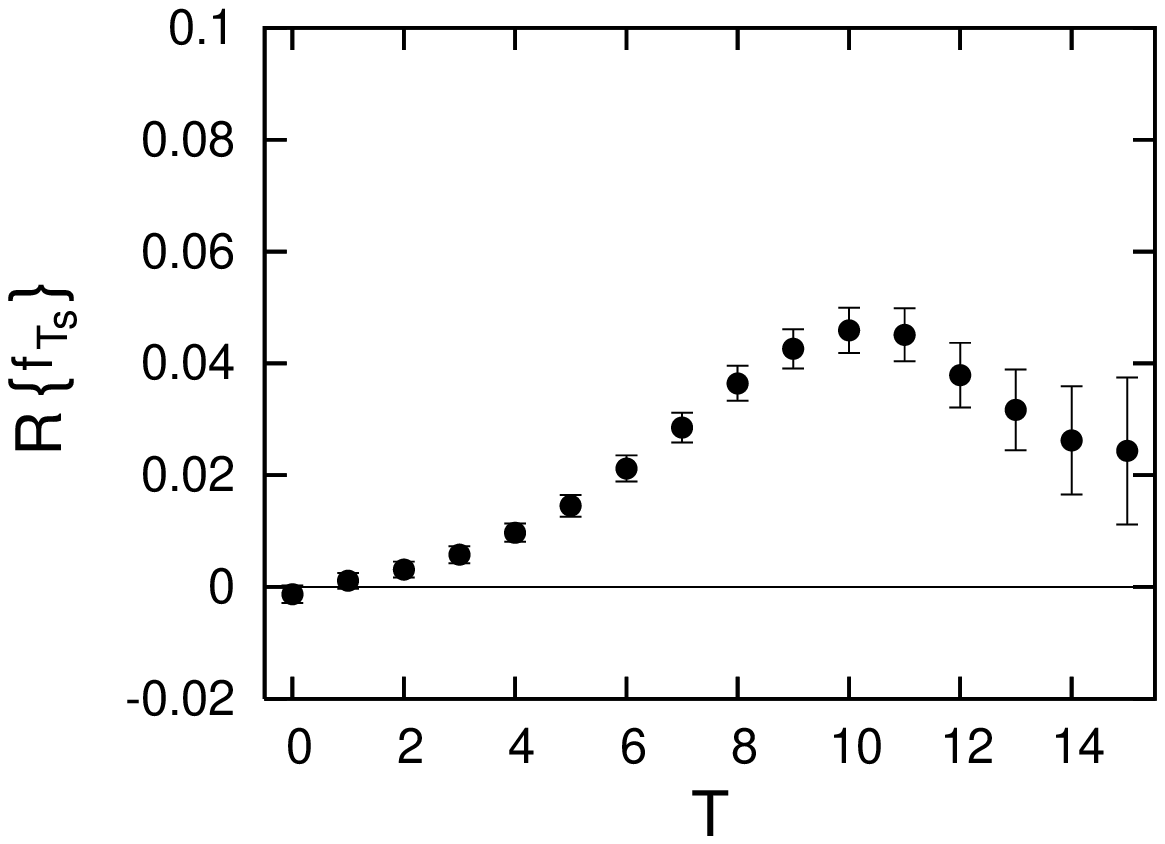}
\vspace{-4.5cm}

\hspace{2cm} {\small $m_{\pi } =356\, \mbox{MeV} $}
\hspace{5.8cm} {\small $m_{\pi } =495\, \mbox{MeV} $}
\vspace{3.7cm}
\caption{Correlator ratio $R\{ f_{T_s } \} $ as a function of sink time $T$,
for the two pion masses considered.}
\vspace{0.09cm}
\label{resfts}
\end{figure}

\begin{figure}
\includegraphics[angle=-90,width=7cm]{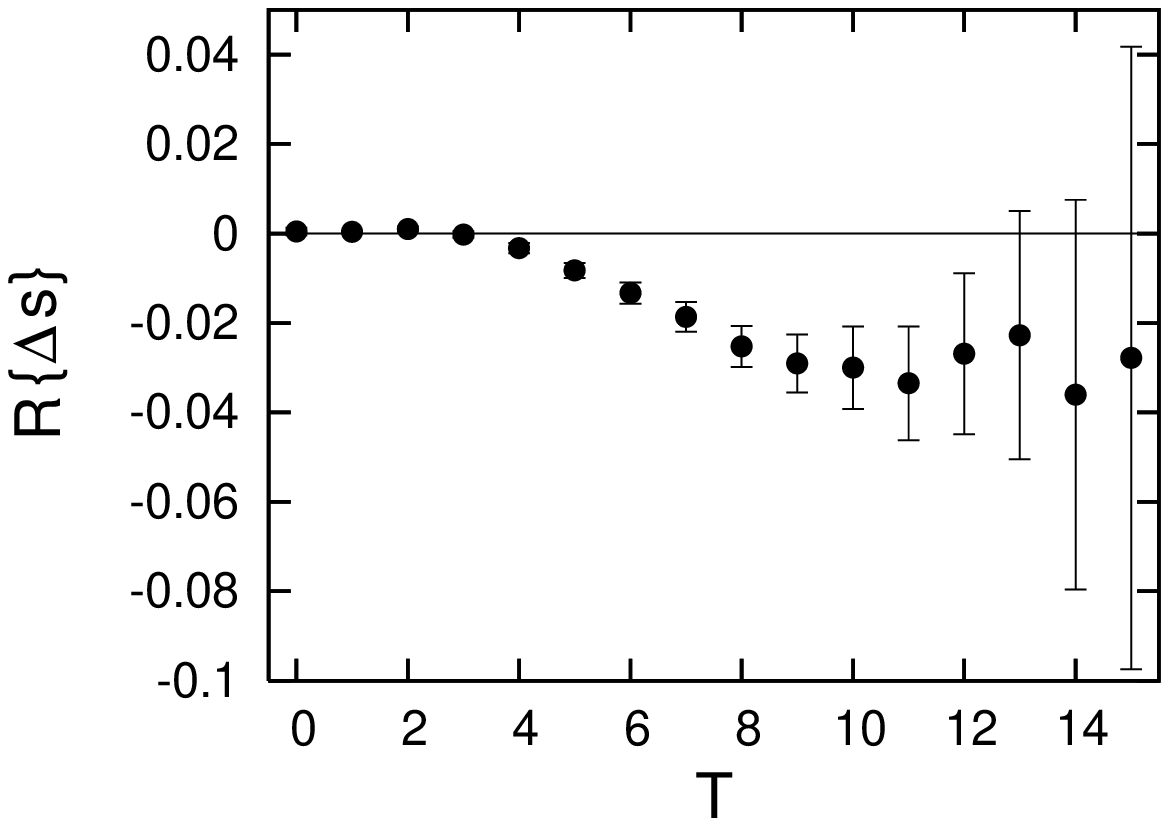}
\hspace{0.95cm}
\includegraphics[angle=-90,width=7cm]{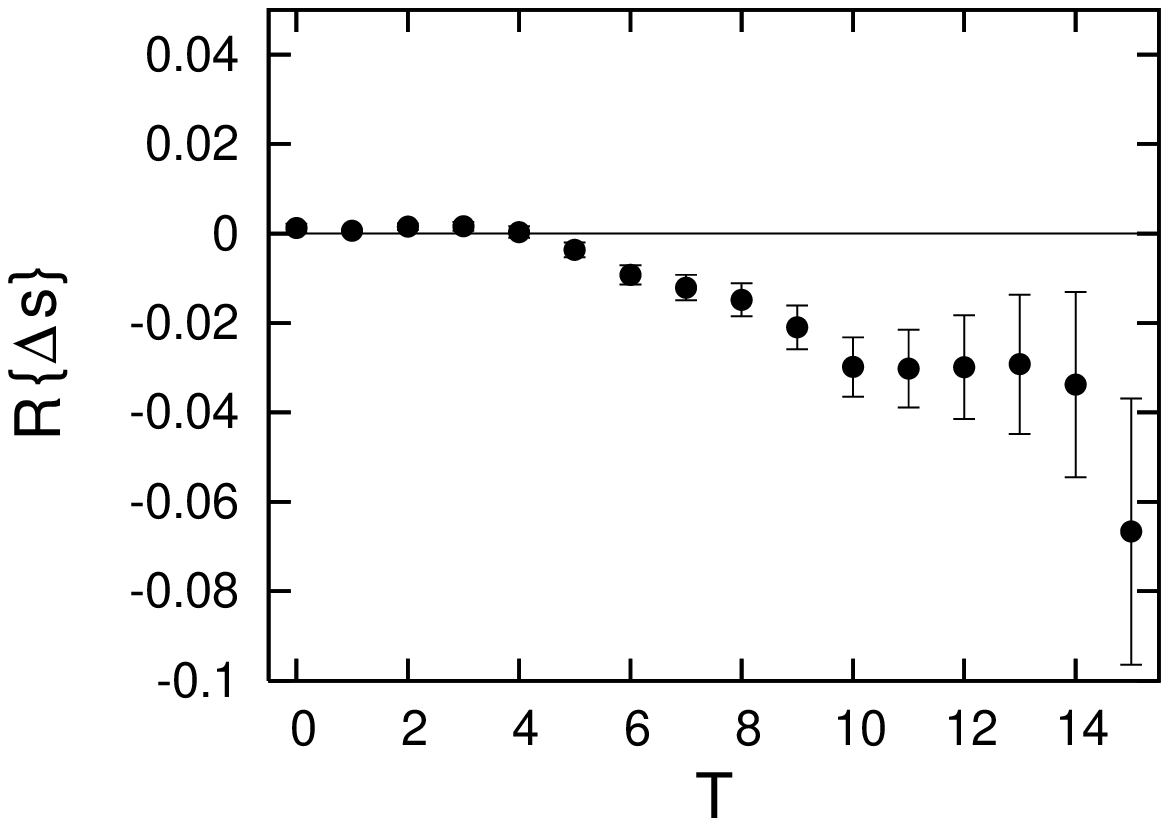}
\vspace{-1.8cm}

\hspace{2cm} {\small $m_{\pi } =356\, \mbox{MeV} $}
\hspace{5.8cm} {\small $m_{\pi } =495\, \mbox{MeV} $}
\vspace{1cm}
\caption{Correlator ratio $R\{ \Delta s \} $ as a function of sink time $T$,
for the two pion masses considered.}
\label{resds}
\end{figure}

Figs.~\ref{resfts} and \ref{resds} display, as a function of sink time $T$,
the correlator ratios $R\{ f_{T_s } \} $ and $R\{ \Delta s \} $,
cf.~(\ref{corrfts}) and (\ref{corrds}), averaged in the fashion described
in the preceding section; for large $T$, these quantities yield the strange
contributions to the nucleon mass and spin, $f_{T_s } $ and $\Delta s$.
Before reaching the physically most relevant region $T\ge \tau $ (where
$\tau $ denotes the temporal location(s) of the operator insertion), the
ratios start out at vanishing values near $T=0$, then gather up magnitude as
the source-sink time interval includes an increasing portion of the stochastic
source region contributing to the quark loop. In the region $T\ge \tau $, the
correlator ratios are expected to level off to approach their asymptotic
value for large sink times. On the the other hand, in this region,
statistical fluctuations become appreciable, and there is therefore
only a short time window in which one can hope to observe this behavior.

In the case of $\Delta s$, cf.~Fig.~\ref{resds}, this expectation is
confirmed rather well. Quantitatively, the estimates of $\Delta s$ obtained
either from the correlator ratio at sink time $T=10$ or from the average
over the correlator ratios at sink times $T=10,\ldots ,14$ yield almost
identical results,
\begin{eqnarray}
\mbox{At } m_{\pi } =356\, \mbox{MeV} : \ \ \ \ \ \ \ \ \ \ \ \ \ \ \ &
\Delta s |_{T=10} = -0.030(9) & \ \ \ \ \ \ \ \ \ \ \ \ \ \ \
\Delta s |_{T=10...14} = -0.030(19) \ \ \ \ \ \ \ \ \ \ \ \ \ \ \ \\
\mbox{At } m_{\pi } =495\, \mbox{MeV} : \ \ \ \ \ \ \ \ \ \ \ \ \ \ \ &
\Delta s |_{T=10} = -0.030(7) & \ \ \ \ \ \ \ \ \ \ \ \ \ \ \
\Delta s |_{T=10...14} = -0.031(11) \ \ \ \ \ \ \ \ \ \ \ \ \ \ \
\end{eqnarray}
where the error estimate in the sink-time averaged case is the jackknife
error extracted by performing the sink time average configuration by
configuration, i.e., the fact that correlator ratios at different
sink times are not independent is taken into account.

On the other hand, in the case of $f_{T_s } $, the behavior of the correlator
ratios as a function of sink time is not as clear-cut. At
$m_{\pi } =356\, \mbox{MeV} $, the correlator ratio for $T>10$ considerably
overshoots the value at $T=10$, whereas at $m_{\pi } =495\, \mbox{MeV} $,
the correlator ratio for $T>10$ decreases again compared to its $T=10$
value. Quantitatively, the comparison analogous to the one performed for
$\Delta s$ above yields
\begin{eqnarray}
\mbox{At } m_{\pi } =356\, \mbox{MeV} : \ \ \ \ \ \ \ \ \ \ \ \ \ \ \ &
f_{T_s } |_{T=10} = 0.043(6) & \ \ \ \ \ \ \ \ \ \ \ \ \ \ \
f_{T_s } |_{T=10...14} = 0.057(11) \ \ \ \ \ \ \ \ \ \ \ \ \ \ \ 
\label{ftsest1} \\
\mbox{At } m_{\pi } =495\, \mbox{MeV} : \ \ \ \ \ \ \ \ \ \ \ \ \ \ \ &
f_{T_s } |_{T=10} = 0.046(4) & \ \ \ \ \ \ \ \ \ \ \ \ \ \ \
f_{T_s}  |_{T=10...14} = 0.037(5) \ \ \ \ \ \ \ \ \ \ \ \ \ \ \
\label{ftsest2}
\end{eqnarray}
The fact that the deviations from the expected plateau behavior occur in
opposite directions for the two pion masses may be an indication that they
are caused by statistical fluctuations; also the error estimates in
(\ref{ftsest1}),(\ref{ftsest2}) are still compatible with this possibility.
In view of the fact that correlator ratios at different sink times are not
independent of each other, the rather smooth behavior of the correlator ratio
in the case of $m_{\pi } =495\, \mbox{MeV} $ does not necessarily contradict
an explanation in terms of statistical fluctuations. Currently, a doubling
of the statistics for the $m_{\pi } =495\, \mbox{MeV} $ case is being
pursued to further explore this issue.

\begin{figure}
\vspace{-0.3cm}
\includegraphics[angle=-90,width=7cm]{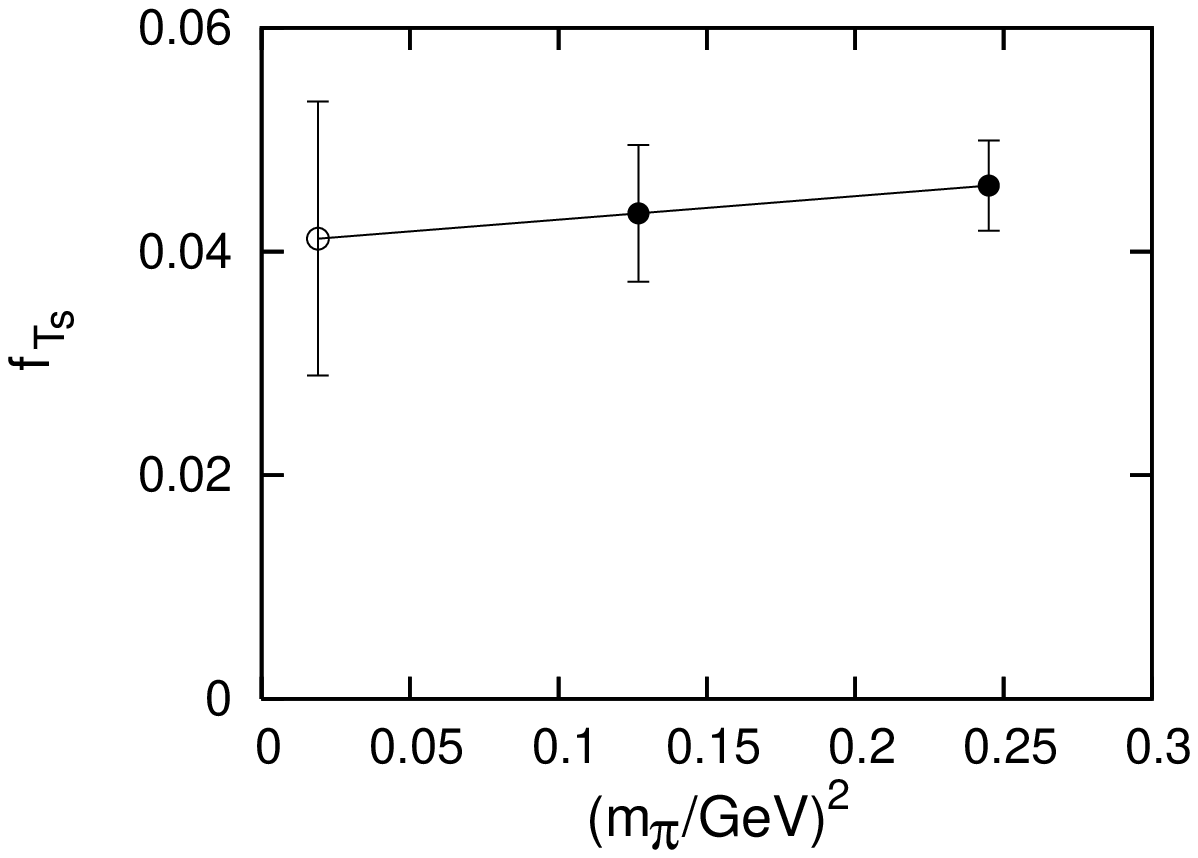}
\hspace{0.95cm}
\includegraphics[angle=-90,width=7cm]{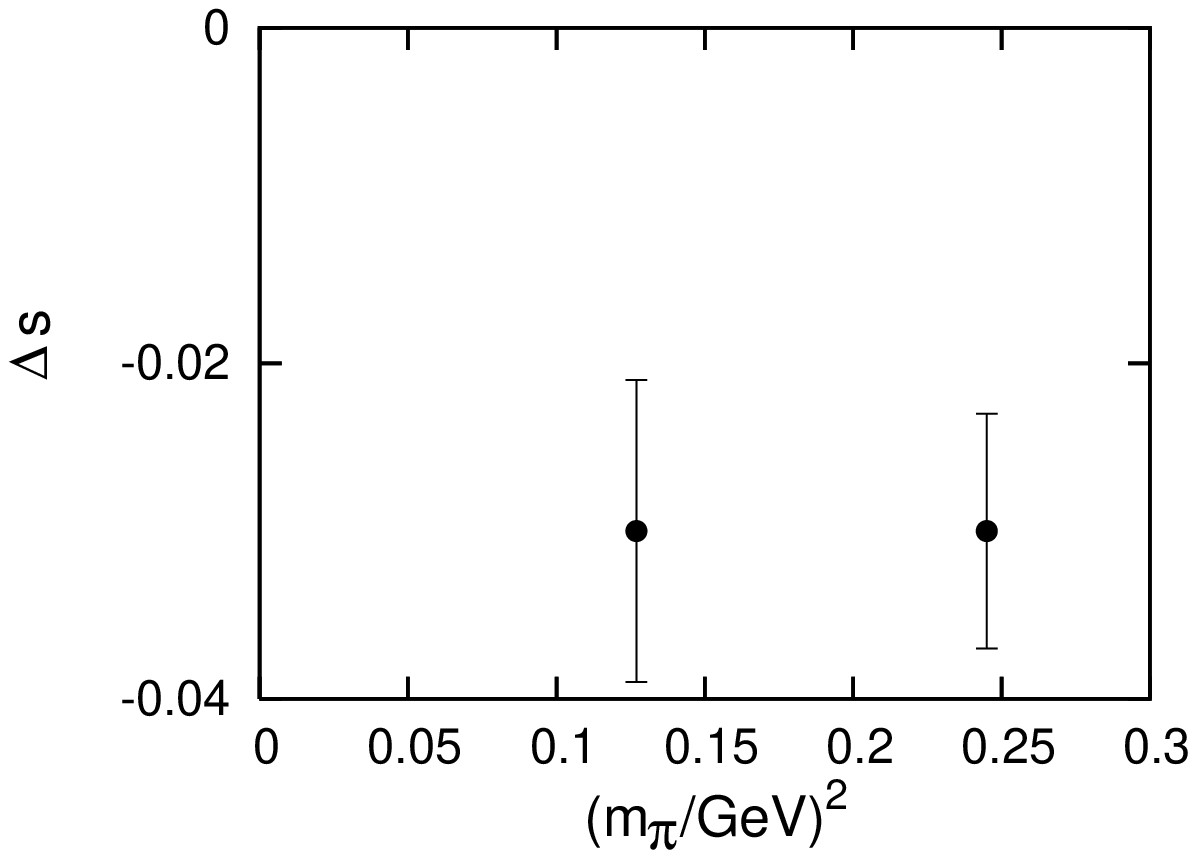}
\vspace{0.1cm}
\caption{Pion mass dependence of the results for $f_{T_s} $ and $\Delta s$.
Filled circles represent measured data; open circle in left-hand panel shows 
chiral extrapolation of the data to the physical pion mass, cf.~main text.}
\label{extrap}
\end{figure}

Fig.~\ref{extrap} summarizes the results obtained as a function of pion
mass, using the correlator ratios at sink time $T=10$ to estimate the
observables; for the case of $f_{T_s} $, the above discussion of the
uncertainties involved in identifying an asymptotic value of the
corresponding correlator ratio should be kept in mind. For $f_{T_s} $,
also a tentative extrapolation to the physical pion mass is displayed;
since the strange scalar matrix element in the nucleon is related via
the Feynman-Hellman theorem to the derivative of the nucleon mass with
respect to the strange quark mass, the chiral behavior follows from
differentiation of nucleon mass formulae obtained in chiral effective
theory \cite{freeman}. The leading dependence is linear in the light
quark mass, i.e., in $m_{\pi }^{2} $; this was used in the extrapolation
shown in the left-hand panel of Fig.~\ref{extrap}. The extrapolated value is
$f_{T_s} =0.041(12)$, corresponding to $m_s \langle N | \bar{s} s |N\rangle
=39(12)\, \mbox{MeV} $.

On the other hand, while no attempt at extrapolating
$\Delta s$ in the right-hand panel of Fig.~\ref{extrap} has been made,
no significant variation with pion mass is seen. It should be noted
that the strange axial current entering the calculation of $\Delta s$
requires renormalization, which has not been included in the results
presented here; however, for the lattice scheme used in this work,
axial current renormalization constants translating to the $\overline{MS} $
scheme at a scale of $2\, \mbox{GeV} $ are consistently very close
to $1.1$, over a wide range of quark masses \cite{lhpc10}. It is therefore
expected that the results for $\Delta s$ only acquire a mild $10\% $
enhancement when translated to the standard $\overline{MS} $ scheme.
Thus, no evidence for unnaturally large strange quark contributions to
nucleon spin is seen in the present calculation.

Of course, it should be noted that no attempt has been made at this
point to quantify several other sources of systematic uncertainty, such
as the dependence on the lattice spacing, lattice size, and, in particular,
the fact that the strange quark mass in the gauge ensembles used lies
appreciably above the physical strange quark mass. Data on the strange
quark mass dependence reported in \cite{freeman}, translated to the
present scheme, indicate that matrix elements such as
$\langle N | \bar{s} s |N\rangle $ acquire a correction amounting to
about $15\% $ (which in $f_{T_s} $ is (over)compensated by the $m_s $
prefactor); it thus appears reasonable to conjecture a $15\% $ systematic
error also for $\Delta s$ from this source, about half the magnitude
of the statistical uncertainty.

\section*{Acknowledgments}
The computations required for this investigation were carried out at the
Encanto computing facility operated by NMCAC, using the Chroma software
suite \cite{chroma}. This work was supported by the U.S.~DOE under grant
DE-FG02-96ER40965.

\end{document}